\documentclass[aps,superscriptaddress,floats,showpacs,floatfix,onecolumn]{revtex4}
\usepackage{bm}
\usepackage{graphicx}
\usepackage{subfigure}
\usepackage{amssymb}
\usepackage{amsmath}
\usepackage{color}
\usepackage[a4paper,left=2.cm, right=2.cm, top=2cm,bottom=2cm]{geometry}
\begin{document}
\newcommand{\joerg}[1]{\textcolor{red}{#1}}
\newcommand{\janet}[1]{\textcolor{blue}{#1}}

\title{ Matching Conditions  and High-Re Anomalies   in Hydrodynamic Turbulence.}
\author{Victor Yakhot}
\affiliation{Department of Mechanical Engineering, Boston University, Boston, MA 02215, USA}
\affiliation{EXA Corporation, 55 Network dr., Burlington, MA, USA 01803, }

\date{\today}

\begin{abstract}
\noindent  
{\bf Direct} transition from low  Reynolds number   "weak"  Gaussian turbulence to    fully developed ``strong''  turbulence   at a  critical  Reynolds number $R^{tr}_{\lambda}\approx 8.91$  has  recently been theoretically predicted   and tested in high resolution   numerical simulations  of  V. Yakhot \&  D. A. Donzis, Phys. Rev. Lett. {\bf 119}, 044501 (2017) \& PhysicaD, {\bf 384-385}, 12 (2018)  
on an  example of a  flow excited by a Gaussian random force.  The matching between  the  low-Reynolds number Gaussian asymptotic ($Re<<Re^{tr}$)  and the multi-scaling one,  dominating the high-Re limit  ($Re>>Re^{tr}$), led to closed approximate  equation for  exponents of moments of derivatives  in a good agreement with experimental data. 
In this paper we  study transition to turbulence   in  Benard (RB) convection  where, depending on the Rayleigh number,  turbulence is produced by  both  weak instabilities of  the  bulk flow  and,  the plume-generating instabilities of the 
  wall boundary layers.  The developed theory explains   
non-monotonic behavior of the  low-Reynolds - number moments of velocity derivatives  $M_{2n}(Re)=\frac{\overline{(\partial_{x}v_{x})^{2n}}}{[\overline{(\partial_{x}v_{x})^{2}}]^{n}}$
observed in  direct numerical simulations of Schumacher et.al  (Phys.Rev.E, {\bf 98},033120 (2018)).  
In the high-Reynolds number limit,  the moments  are given by $M_{2n}\propto Re^{\rho_{2n}}$  with the   exponents $\rho_{2n}$ slightly different from those  in a Gaussian-stirring  case of Refs. [3]-[4].  This  may be related to  universality classes defined by  production mechanisms.

 
\end{abstract}
\maketitle

\section{Introduction}

 \noindent  
 Transition from laminar   to turbulent flow  was  discovered and analyzed by Osborn Reynolds in 1883, who   reported  emergence  of "sinuous" motions    out of  a direct and steady  water flow  in a pipe.  Moreover, Reynolds quantified the phenomenon in terms 
 of  dimensionless parameter $Re=UL/\nu$,  later called Reynolds number.  Here $U$ and $L$ denote  mean velocity across the pipe of radius $L$. In this  work, Reynolds introduced a critical parameter $Re=Re_{cr}$,   so that at $Re\leq Re_{cr}$ the flow was laminar, with steady parabolic velocity profile $U(r)$.  He noticed the appearance of  irregular or random fluctuations  ${\bf v}({\bf x},t)$  when   $Re\geq Re_{cr}$. With increase of $Re>Re_{cr}$,  the amplitude and degree of randomness increased which made analysis  of the flow very hard. Interestingly, Reynolds was the first to suggest description of this  flow using statistical methods.
  To this day, the question of  structure and statistics of velocity fluctuations ${\bf v}({\bf x,t)}$ as a function of $ Re-Re_{cr}\rightarrow \infty$
  remains open.\\
  \noindent   Depending on geometry and physical mechanisms, various laminar flows become unstable at  widely different Reynolds numbers $Re=VL/\nu$, where $V$ and $L$  are characteristic velocity and length scale  of a flow.    One can introduce  dimensionless critical number  
 $Re_{cr}$ marking first instability of a laminar flow pattern.  As  $Re-Re_{cr} \rightarrow 0+$,   low - intensity  
 velocity fluctuations   are described as,  usually  Gaussian,  random field, which can loosely be called  ``weak 
   or soft turbulence''.    Some qualitative ideas   can be obtained  from Landau theory considering a  {\it stationary}  flow ${\bf v}_{0}(\bf x)$  with a small time-dependent perturbation ${\bf v}_{1}({\bf x},t)=A(t){\bf f}(\bf x)\propto f({\bf x})e^{\gamma t}e^{-i\omega_{1}t}$ where $\omega_{1}\gg |\gamma|$.   In the vicinity of  a transition point , where $\gamma\propto Re-Re_{cr}\rightarrow 0$,  one can write
   
 \begin{eqnarray}
 \frac{d|A|^{2}}{d t}=2(Re-Re_{cr})|A|^{2}-\alpha  |A|^{4}
 \end{eqnarray}
 
 \noindent When $\gamma  \approx Re-Re_{cr}>0$, the  growing  with time amplitude $A(t)$    saturates at $|A|_{max}\propto \sqrt{(Re-Re_{cr})/\alpha}$.
 Extrapolating this into interval $Re>>Re_{cr}$,  we obtain $|A|_{max}\propto \sqrt{Re}$.  This result can  numerically  be sufficiently accurate for practical purposes  when $Re-Re_{cr}>0$ is finite   though small enough for the $O(A^{6})$  contributions to (1) be neglected.       The important feature of (1) is that  no randomness is present in Landau's  theory which assumes  that   equation (1) is an outcome of  averaging  over high -frequency  phases with  $\omega_{1}>>\gamma=O(Re-Re_{cr})\rightarrow 0$.

\noindent  Landau  assumed that with further increase of the Reynolds number,  the field ${\bf v}_{0}+{\bf v}_{1}$ becomes unstable
 i.e. its perturbation ${\bf v}_{2}({\bf x},t)$ grows into a periodic flow with frequency $\gamma_{2}\approx 2\gamma$ and so on.  
 While this theory is physically appealing, its main 
 drawback  is  the fact that the Reynolds number of   the ``second'' instability  
 generating  small-scale fluctuations  is unknown 
 and it is not clear how one can calculate it when $|A| $is not small.  Various  attempts to treat (1) as a first two terms of the Taylor expansion by  adding a few high-order powers in $A$ led to unsurmountable complications [2]. 
 
\noindent The passage  to strong turbulence involves a few steps : a.\ laminar or regular low - Reynolds number field ${\bf U}({\bf x},t)$  which is a solution to the Navier-Stokes equations b. \  theoretical or experimental understanding  of its stability;  c.\ study of fluctuations and their interactions with each other  and with a mean flow.  Each step of this  program is extremely involved and difficult due to in general complex  geometry and  lack of a small parameter.  Not surprisingly,  the strong turbulence problem   is a subject of more than a century of experimental and theoretical efforts.  In this paper we are interested in a completely different kind of transition to fully developed ``strong'' turbulence not involving instability of a laminar, regular,  velocity field ${\bf v}_{0}(\bf x)$.\\

\noindent 
{\bf ``Reynolds numbers''  in a fully developed turbulent flow.}
The Reynolds' description of  transition to turbulence was based  on  dimensionless coupling constant  constructed from characteristic velocity $V$  and length-scale $L$ of a laminar background flow. 
 It was realized later that  the  Reynolds number based on Taylor scale $\lambda$  and rms velocity $v_{rms}=\sqrt{\overline{v^{2}}}$ was  a better  descriptor of a stochastic flow characterized, for example, by   "structure functions"  

$$S_{n}=\overline{(v_{x}({\bf x})-v_{x}({\bf x}+r{\bf i}))^{n}}\propto  (\frac{r}{L})^{\zeta_{n}}$$

\noindent where $v_{x}$ is the $x$-component of velocity field and ${\bf i}$  is the unit - vector in the $x$-direction. The moments of  derivatives,    including   those of dissipation rate,  ${\cal E}=\nu(\frac{\partial v_{i}}{\partial x_{j}})^{2}$,   
we are  interested in this paper are defined as:

$$M_{2n}=\frac{\overline{(\partial_{x}v_{x})^{2n}}}{{\overline{(\partial_{x}v_{x})^{2}}^{n}}}\propto Re^{\rho_{2n}}$$

\noindent where the large-scale Reynolds number $Re$ is defined in Table 1. It became clear that the so-called Kolmogorov's scaling $\zeta_{n}=n/3$ and $\rho_{2n}=n$  is not valid for  $n\neq 3$ and the moments of orders $m$ and $n$ with $m\neq n$ 
are given by some "strange" numbers not  
related to each other by dimensional considerations.  This feature of strong turbulence, called "anomalous scaling", is the signature of strong interactions between modes  in non-linear systems..
For many years theoretical evaluation of anomalous 
 exponents $\zeta_{n}$ and $\rho_{n}$ was considered one of the main goals   of the proverbial  "turbulence problem". 
 It was shown both theoretically and numerically  in Refs.[3]-[4] that possible reason for this difficulty is hidden in the fact that each moment $S_{n}(r)$  and  $M_{n}$ should be characterized by its "own" $n$-dependent Reynolds number $\hat{R}e_{n}$ based on characteristic velocity $ \hat{v}(n,n)$, defined in Table 1, and  that a widely used parameter $v_{rms}=\hat{v}(2,2)$ is simply one of an infinite number of characteristic velocities describing turbulent flow. 
 The multitude of  dynamically relevant Reynolds numbers,  necessary for description of turbulence,  is  defined in Table 1.

\begin{table}[h]
\begin{tabular}{l|l}
\hline
Reynolds number & Description \\ \hline \hline
$v_{rms}=\sqrt{\overline{v^{2}}}$ & root-mean-square velocity\\
$\hat{v}(m,n)=\overline{|v|^{m}}^{\frac{1}{n}}$ &moment of order $m/n$;  $v_{rms}=\hat{v}(2,2)\equiv \hat{v}_{2}$\\
$Re =v_{rms} L/\nu$ & large-scale Reynolds number\\
$\hat{Re}_{n}=\hat{v}(n,n)L/\nu$ & Reynolds number  of the   $n^{th}$ moment \\
 $R_\lambda = v_{rms} \lambda/\nu$ & Taylor Reynolds number;
                          $\lambda=15\nu u_{rms}^2/{\cal E}$  \\
$Re_{n}^{tr}$ & transition point 
                     for moments of order $n$ \\
$\hat{R}e_n=\hat{v}(n,n)L/\nu$ &  
probes regions with different amplitudes of velocity gradients\\
$\hat{R}_{\lambda,n}=(5L^4/3{\cal E} \nu)^{1/2}\hat{v}(2n,n)$ &  
 order-dependent Taylor-scale Reynolds number\\
\\
\hline
\end{tabular}
\label{tab:res}
\caption{Summary of Reynolds numbers used in this work.}
\end{table}

\subsection{Matching condition and anomalous exponents.  Application to direct transition.} 
 
\noindent In Landau's theory of "laminar-to-turbulent transition", the Reynolds number is defined on a ``typical'' characteristic  velocity $V$ and length-scale $L$ depending   on flow geometry, dimensionality, 
physical mechanisms responsible for  instability and other factors characterizing large-scale ordered (laminar) flow.  Therefore, in this  approach $Re_{cr}$  varies in an extremely  wide range of parameter variation. 
To study  dynamics of velocity fluctuations  it is useful to define the   Reynolds number   $Re=v_{rms}L/\nu=\sqrt{\overline{v^{2}}}L/\nu$ based entirely on fluctuating velocity  ${\bf v}$ for which $\overline{\bf v}=0$.  
To avoid difficulties  related to instabilities of a laminar flow,   we studied  the dynamics governed by  the  Navier-Stokes equations  in an infinite fluid stirred by a Gaussian random forcing  acting on a finite scale $r\approx L$  Refs. [3]-[4]:

\begin{equation}\partial_{t}{\bf v}+{\bf v\cdot\nabla v}=-\nabla p +\nu\nabla^{2}{\bf v} + {\bf f} \end{equation}

\noindent   $\nabla\cdot {\bf v}=0$.  Here the density is taken $\rho=1$ without loss of generality.  A random Gaussian noise ${\bf f} $ is  defined by  correlation function:

 \begin{equation}\overline{f_{i}({\bf  k},\omega)f_{j}({\bf k'},\omega')}= (2\pi)^{d+1}D_{0}(k)P_{ij}({\bf k})\delta({\hat{k}+\hat{k}'})\end{equation}

\noindent  where the four-vector $\hat{k}=({\bf k},\omega )$ and projection operator is: $P_{ij}({\bf k})=\delta_{ij}-\frac{k_{i}k_{j}}{k^{2}}$.    It is clear from (2)-(3) that in the limit $D_{0}\rightarrow 0$ the nonlinearity is small and ${\bf v}(\hat{k})\approx G^{0}{\bf f }=O(\sqrt{D_{0}})$,  where the ``bare'' Green function is  $G^{0}=1/(-i\omega +\nu k^{2})$. In this limit the velocity field is Gaussian with  the derivative moments $M_{2n}=\overline{(\partial_{x}v_{x})^{2n}}/\overline{(\partial_{x}v_{x})^{2}}^{n}\approx (2n-1)!!$. \\

\noindent  
\noindent {\bf As stated above,  we consider an infinite fluid stirred at a finite scale $L$. This means that if  linear dimension of a fluid is ${\cal L}\rightarrow \infty$, then the flow is generated by $N={\cal L}^{3}/L^{3}\rightarrow \infty $ random, uncorrelated,  stirrers, each one defining a statistical realization. Therefore, one can  describe a  flow either in terms of local parameter fluctuations or, equivalently, by statistical ensemble with corresponding  probability densities (PDFs) . This will be demonstrated  in detail below.}\\ 



\noindent  Due to  the lack of small expansion parameter,  all  renormalized perturbation theories   applied to  the problem (2)-(3),  failed to yield  experimentally 
observed  anomalous scaling of velocity increments and derivatives.  This failure is easily 
explained in terms of a single dimensionless (``dressed'' ) coupling constant  $Re_{T}\approx v_{rms}L/\nu_{T}=O(1)$ 
appearing in perturbation expansions.  Here, $\nu_{T}$ is effective (turbulent) viscosity accounting for   interaction of large - scale "eddies" on a scale $r\approx L$  with small-scale velocity fluctuations [1].  It has been shown in [3]-[4] that  describing 
multi-scaling processes,  one has to  introduce an  infinite number of different coupling constants reflecting the multitude of  scaling exponents. 

\noindent We  can  seek a non-perturbative  solution  satisfying  two asymptotic  constraints:   
{\bf in the ``weak turbulence''  range  $Re\ll Re_{2n}^{tr}$  ($D_{0}\rightarrow 0$), the Gaussian solution  $M_{2n}=(2n - 1)!!$ follows directly  from (2)-(3).  In the opposite strongly  non-linear limit   $Re\gg Re_{2n}^{tr}$, the   moments   $M_{2n}=\Gamma(Re_{2n},2n)\approx A_{2n}Re^{\rho_{2n}}$  with not yet known  amplitudes   $A_{2n}$ and exponents $\rho_{2n}$. 
The two limiting curves match  at  the $n$-dependent transitional Reynolds numbers  $Re_{2n}^{tr}$,  investigated in detail in Refs.[3]-[4]. } Thus, 
at a transition point  of the $2n^{th}$ moment  $Re=Re_{2n}^{tr}$:

\begin{eqnarray}
M_{2n}=(2n-1)!!\approx A_{2n}(Re_{2n}^{tr})^{\rho_{2n}}
\end{eqnarray}

\noindent   On Fig.1,  these  ideas have  been confirmed  by direct numerical simulations (DNS) of the 
 the moments of  derivatives $M_{2n}$ vs Reynolds number based on the Taylor scale 
 $R_{\lambda}=\sqrt{\frac{5}{3{\cal E}\nu}}v_{rms}^{2}$.   We can  see  horizontal lines corresponding to the Re-independent normalized Gaussian 
moments $M_{2n}=(2n-1)!! $ for  $2 \leq n \leq 6$.   

\begin{figure*} 
\includegraphics[height = 4cm]{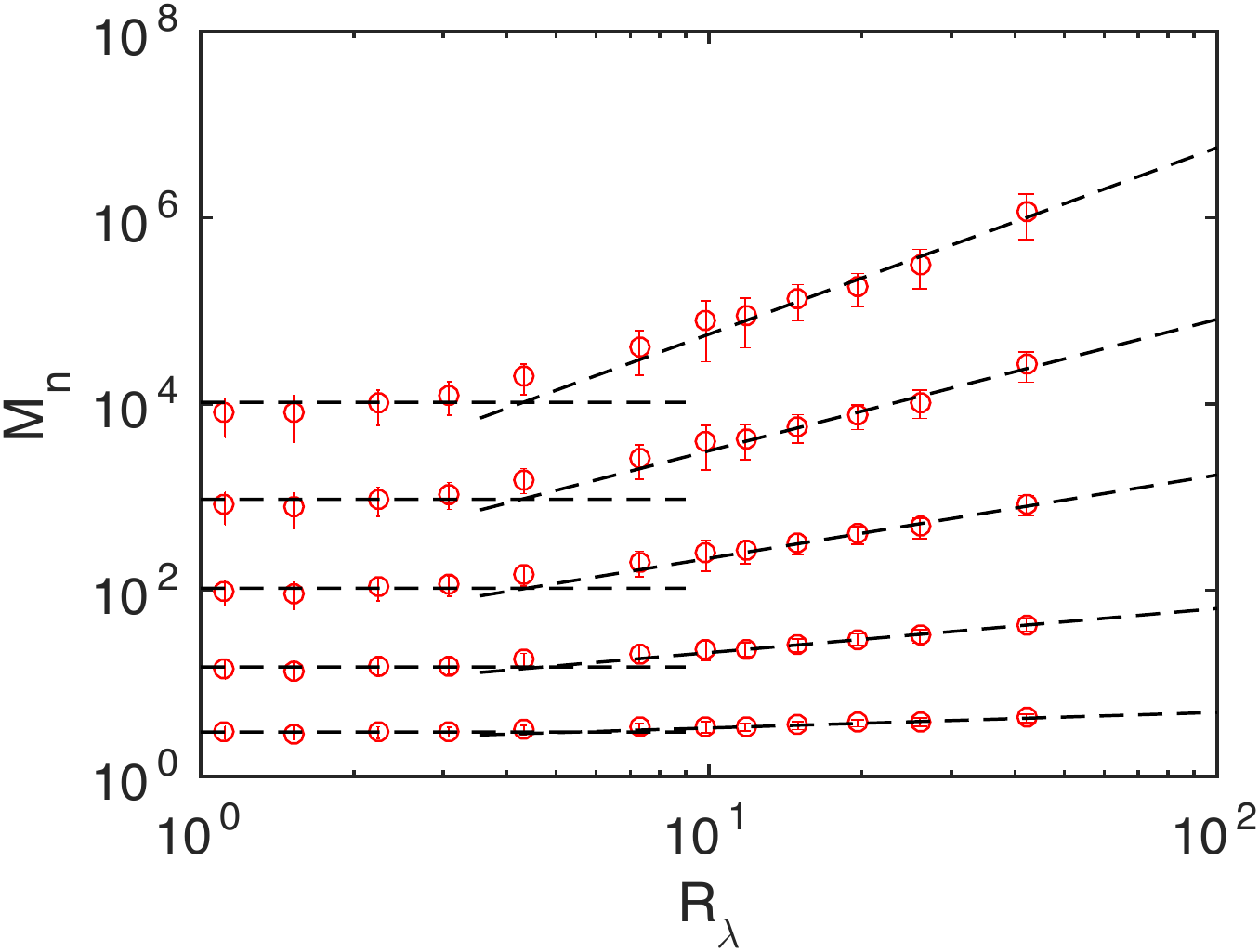}
\includegraphics[height=4cm]{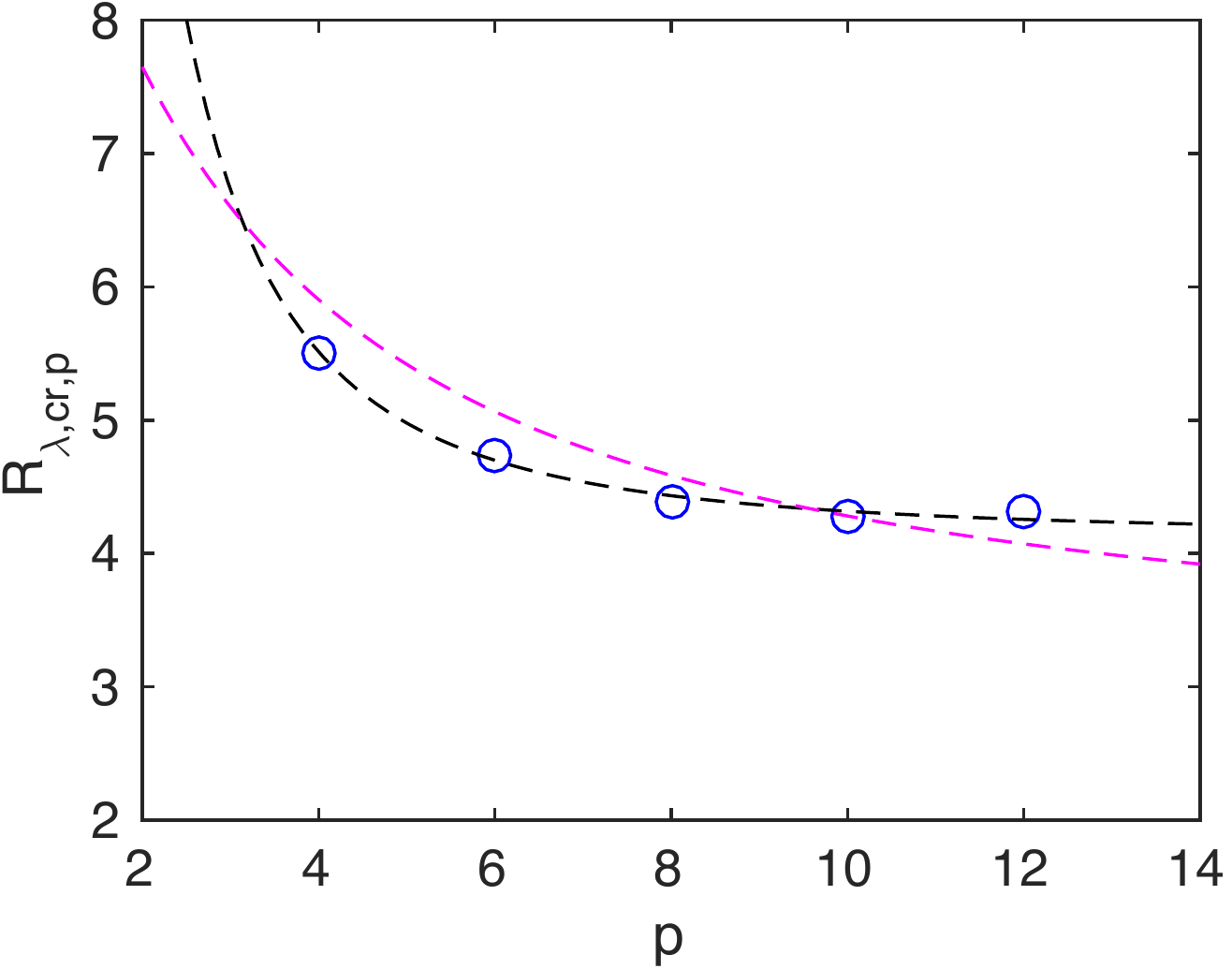}
\includegraphics[height =4cm]{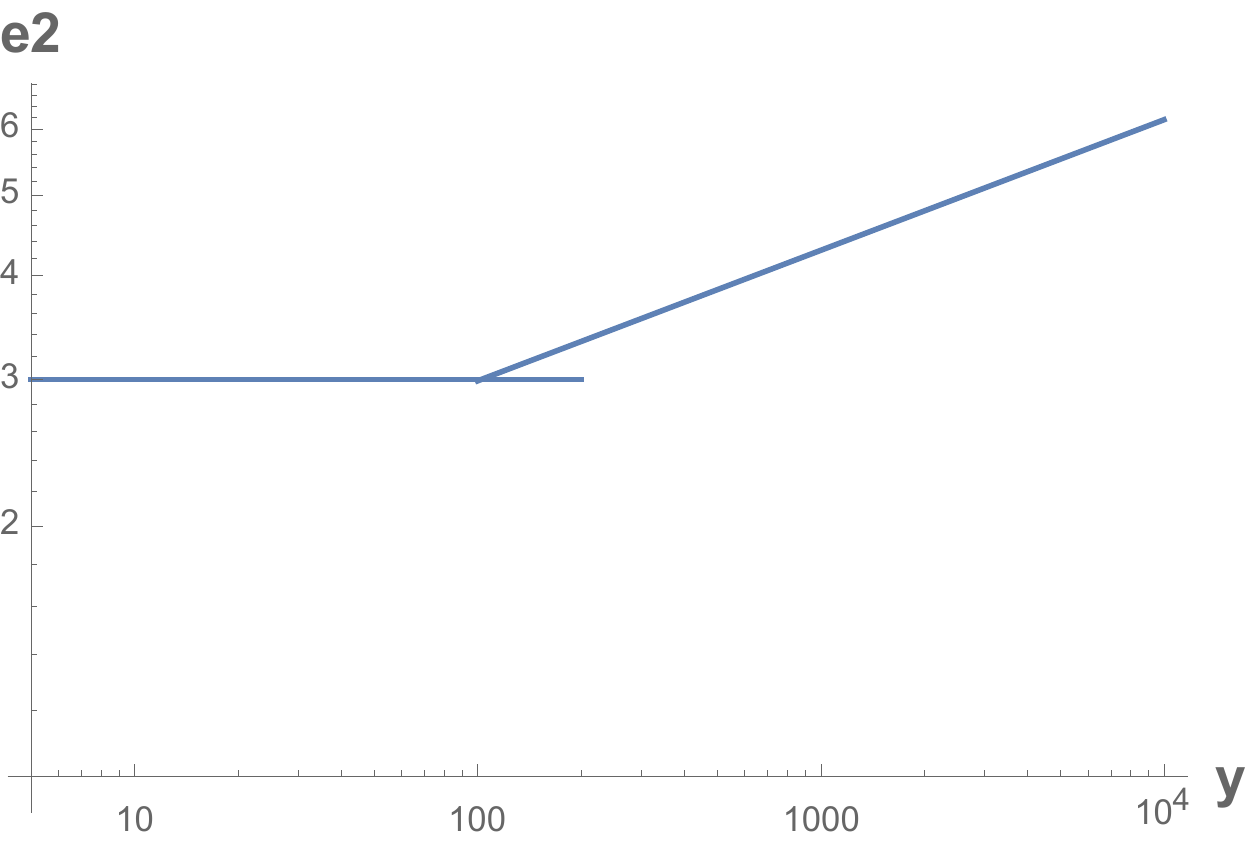}
\caption{ Left   panel:  Normalized moments of velocity gradients $M_{2n}$ from direct numerical  solutions to the Navier-Stokes equations (2)-(3).  Both  asymptotics,   leading to  predicted matching relation  (4),  are clearly seen. Middle panel: Reynolds number dependence of different-order  moments $M_{2n}(Re)$.   From Ref.~[3]-[4].  Right panel: theoretical construction leading to the second-order moment $e_{2}\propto \overline{{\cal E}^{2}}$ in the entire range of $Re$-variation.  The low-Reynolds 
 input : $Re_{2}^{tr}\approx 100-120$ and $e_{2}=3.0$  for $Re<100$.  This gives $d_{2}\approx 0.157-0.167$. In the strong coupling range $e_{2}=1.45\times Re^{0.157}$.}
\end{figure*}

\noindent  We would like to stress an important point: $v_{rms}$  characterizes   typical or  relatively mild velocity fluctuations. In general, to be able to predict   rare,  extreme,  events  we   introduce   $\hat{v}_{2n}=L^{2}\overline{(\partial_{x}v_{x})^{2n}}^{\frac{1}{n}}\propto A_{2n}^{\frac{1}{n}}Re^{\frac{\rho_{2n}}{n}}$ and $\hat{R}_{\lambda, n}^{tr}=\sqrt{\frac{5}{{3\cal E}\nu}}\hat{v}_{2n}\approx 8.91$ derived in Refs.[3]-[8].  To calculate  large- scale transitional Reynolds number  we introduce velocity scale $v_{0}=v_{rms}$  so that $Re=v_{0}L/\nu$ and :

$$\hat{R}_{\lambda,n}^{tr}=\sqrt{\frac{5}{3{\cal E}\nu}}\hat{v}_{2n}=A_{2n}^{\frac{1}{n}}(Re^{tr})^{\frac{\rho_{2n}}{n}+\frac{1}{2}}\approx 8.91$$

\noindent It follows from this relation that transition to strong turbulence in different realizations or different-order-moments occurs at a constant $R^{tr}_{\lambda,n}=8.91$  but at {\bf different}  $Re^{tr}=v_{rms}L/\nu$ based on the r.m.s. velocity coming from the second-order moment. This result, theoretically evaluated in [5]-[8],  is consistent with the empirical   ${\cal K}-{\cal E}$ model 
giving the large-scale ``dressed'' viscosity $\nu_{T}=0.0845{\cal K}^{2}/{\cal E}$,   used in  engineering simulations during last fifty years [9]. Indeed: with ${\cal K}=v_{rms}^{2}/2$

$$R^{tr}_{\lambda}\equiv R^{tr}_{\lambda,2}=\sqrt{\frac{5}{3{\cal E}\nu_{T}}}2{\cal K}\approx 8.88$$

\noindent  and 

$$Re_{n}^{tr}=[\frac{\hat{R}_{\lambda,n}^{tr}}{A_{2n}^{\frac{1}{n}}}]^{\frac{2\rho_{2n}}{2\rho_{2n}+n}}$$

The somewhat  ``unexpected'' but qualitatively reasonable  consequence of this result,  is  seen on Fig.1,  where the  onsets  of anomalous 
scaling for different moments $M_{n}$  are observed at very different $Re_{n}^{tr}$ but at a  {\bf single} $n$-independent  $\hat{R}_{\lambda,n}^{tr}\approx 9.0-10$.  For large enough $n$,  $A_{2n}^{\frac{1}{n}}$ is a weakly dependent function of $n$ which can be calculated from the $Re _{2}^{tr}\approx 9-10$. 
Thus, one can easily express    $Re_{n}^{tr}$ in terms of $\hat{R}_{\lambda,n}^{tr}\approx 9.0-10$ [3] - [4] and close the equation (4) for $\rho_{2n}$.  The results are presented in Table II  and compared with the data on the middle panel of Fig.1.


\subsection{Matching condition: numerical procedure for high-Reynolds number limit.}

 \noindent   In addition to  the   ``classic'' problem of    anomalous exponents $d_{n}$ and $\rho_{n}$, the study of Refs.[3] -[4]  opened  up a new question of possible universality of transitional Reynolds number $\hat{R}^{tr}_{\lambda,n}=\sqrt{\frac{5}{3{\cal E}\nu}}\hat{v}_{2n} \approx 8.91$ derived  in the Renormalization Group analysis of turbulence in the limit $r\rightarrow L$ [5]-[9]. 
 The  possible universality of this result may have  important consequences for  numerical simulations demonstrated in Fig.1 where the analytic theory  is  compared to  the low Reynolds number DNS   on the two left panels.  In the  Gaussian  forcing case [3]-[4]:
 
 \begin{eqnarray}
 Re^{tr}_{n}=C(\hat {R}^{tr}_{\lambda,n})^{\frac{n}{d_{n}+\frac{3n}{2}}}
 \end{eqnarray}
 
\noindent and  at transition points the matching  condition must be satisfied:

\begin{eqnarray} 
e_{n}= \overline{({\cal E}/\overline{\cal E})^{n}}= (2n-1)!!=C^{d_{n}}(\hat {R}^{tr}_{\lambda,n})^{\frac{nd_{n}}{d_{n}+\frac{3n}{2}}}
\end{eqnarray}
 
 \noindent  where  $\hat{R}_{\lambda,2n}^{tr}\approx 8.91$ independent on $n$.  One can easily derive a simple estimate $Re \approx  1.5 R_{\lambda,2}^{2}$ giving $Re_{2}^{tr} \approx 100-200$  resulting in $C\approx 100-200$.  This closes the equation  for exponents $d_{n}$ and $\rho_{2n}$: if, as in the problem $(2)-(3)$,  $\overline { \cal E}\approx D_{0}=O(1)$, then $ \rho_{2n}=d_{n}+n$. The details are presented in  [4]. 
 \noindent The possible universality of transitional $R^{tr}_{\lambda,n}$  enables   high-Reynolds number computations  of flows based on the low-Reynolds number data obtained either theoretically or numerically.
 The  matching procedure is qualitatively demonstrated on the right- most panel of Fig.1 on an example of the   moment $e_{2}(y)$ where $y\propto Re_{2}-Re_{2}^{tr}$. It consists of three  main steps: a. calculate or compute  the moments of derivatives in the linear low-Reynolds number limit $\hat{R}e _{n}\leq \hat{R}e_{n}^{tr}\approx  120$  or $\hat{R}_{\lambda,n}\leq R_{\lambda,n}^{tr}\approx 8.91$.   b. This allows evaluation of the exponents $d_{n}$ and $\rho_{n}$. c. Extrapolation of  an  assumed high-Reynolds number solution 
  $e_{n}=C^{d_{n}}Re^{d_{n}}$ back to the transition point, $Re\rightarrow Re^{tr}_{n}$.    
  c. Plot the resulting dependence in  the entire range $Re\geq Re_{n}^{tr}$. 
 Below we  generalize this scheme to a much  more complex system.
 
 \begin{table}
\begin{ruledtabular}
\begin{tabular}{ccccccccccc}
\hline
$n$ & $1 $ &  $ 2 $ & $3$ & $4$ \\
$R_{\lambda,2n}^{tr}$  & $8.91$ & $5.5$ &  $4.8$ & $4.5$  \\
$Re_{2n}^{tr}$ & $126$ & $45$ & $35 $ & $ 30 $\\
 \end{tabular}
 \end{ruledtabular}
\caption{ Transitional Reynolds numbers based on Taylor scale $R^{tr}_{\lambda,2n}=\sqrt{\frac{5}{3{\cal E}\nu}}v_{rms}^{2}$  of the moments $M_{2n}$. 
With $\hat{v}_{2}=\overline{v^{2n}}^{\frac{1}{n}}$, the  modified Reynolds number $\hat{R}_{\lambda,2n}^{tr}=8.91$ is  independent on $n$.}
\end{table}
 

 \begin{table}
\begin{ruledtabular}
\begin{tabular}{ccccccccccc}
\hline
 $ \rho_{n} $ & $EXP$ &$ GAU$& $DNS$\\
\hline
$\rho_{1} $&  $0.48 $& $0.46$& $0.455$\\
$ \rho_{3}$ &$1.55$& $1.58$ & $1.478$\\
$ \rho_{4}$ & $2.12 $& $2.19$ & $2.05$\\
 $ \rho_{5}$& $2.7$&  $2.82 $& $2.66\pm 0.14$\\
 $ \rho_{7} $& $ 3.92 $& $4.13$ & $3.99\pm 0.65$\\
 \end{tabular}
 \end{ruledtabular}
\caption{Comparison of exponents $\rho_{2n}=d_{n}+n$  with the outcome of numerical  simulations (DNS)  and  Theory.   $EXP$ and $GAU$ from  expression (14)  with  the moments $e_{n}=n! $  and $e_{n}=(2n-1)!!$ in  flows  stirred by exponential and Gaussian  random forces, respectively.     }
\end{table}


 \section{  Boundary layer effects. }
 
 \noindent The simplified problem of Refs. [3]-[4], described above, dealt with an artificial situation of 
   {\bf direct}  transition between a well-defined    Gaussian  state of a  fluid  and the  non-linearity-dominated strong turbulence. 
   In the case of  "direct transition", described by (2)-(3),    turbulence  is produced  by a singe physical mechanism, i.e. external random forcing. 
 In real-life flows  various   randomness - generating mechanisms often act   simultaneously: for example  in wall flows 
 turbulence is   generated by  instability of a quasi-laminar flow pattern in the bulk and by instability of  viscous wall layers generating   powerful bursts reaching  bulk of  a flow. 
Therefore,  while as $y\propto Re-Re_{cr}\rightarrow 0+$ ,  the velocity field   often obeys  Gaussian statistics, 
at  intermediate, but still  linear regime,   due to  the wall boundary layer instability, 
transition to strong turbulence and anomalous scaling  of velocity derivatives may happen not from the Gaussian state.
Below we address this problem. \\

\noindent The  problem of thermal convection  in a fluid heated from below is  a remarkable  laboratory for  studying different areas of physics like  heat  conduction,   pattern  formation,  their stability and instabilities as well as transitions to chaos and  strong turbulence.   It is perfectly  suited  for studies  of small-scale  structure in a strongly non-linear  turbulent state in the limit $Re\rightarrow \infty$. 
In general, the problem is very hard, for it involves  the  first   instability leading to  rolls,  generation of  the low-Re ``weak turbulence''  which is  a precursor to the strong  ``hard'' turbulence we are interested in this paper. The number of both experimental and theoretical publications dealing with  RB convection published in the last few decades is enormous and it is impossible  even to briefly  review  them.   Majority of the work in the field dealt with the large-scale global properties of the phenomenon leading to predictions of  the heat transfer as a function  of various large-scale parameters.  Here we are interested in the small-scale velocity and velocity derivatives fluctuations, which is a relatively new and interesting topic.   

\noindent This problem has been addressed in the DNS published in a recent paper [10]  on  the RB convection  with Prandtl and Reynolds  numbers varying in the range $0.005\leq Pr\leq 100$ and $1\leq Re\leq 1000$.  At large  $Re\approx 100-2000$ the scaling exponents of the  first two moments  of kinetic energy  dissipation rate  were  similar to those  observed in Ref.[4], indicating  possible universality. On the other hand, the  low-Re behavior of a flow, reflecting some structural transitions,  was much more complex and 
appearance of anomalous scaling at $Re\approx 100$ was definitely not from a Gaussian state of Refs.[3]-[4].  In  this case,  unlike  the direct transition,  the Reynolds number dependence of moments of derivatives $M_{2n}(Re)$ was nonmonotonic  having a well - pronounced minima in the low-Re interval  [10]. 
While this paper shed light on many important phenomena related to the Prandtl number dependence of the heat transfer,  the details of  statistics of the dissipation rate fluctuations, including the non-monotonic behavior of the moments, remained somewhat unresolved, mainly due to large  difference between thermal and viscous boundary layers  substantially complicating 
 the situation.    \\
 
 Below,  based on a general approach developed by Sinai  et.al. [10]-[12]  we consider a greatly 
 simplified problem of thermal convection in a gap $H$  between two infinite plates.

\subsection{Phenomenology.}

In this paper we are interested in the  small-scale behavior of a flow between two infinite  plates separated  by the gap $H$. The low plate  at $z=-H//2$ is heated by an electric current $I$.  Due to the energy conservation, the heat flux averaged over horizontal planes $J(z)=cost$ and we keep the top and bottom plates under  constant temperature difference $\Delta$. 
We consider  the coupled three-dimensional equations of motion for velocity and  temperature fluctuations $v_{i}$  and $T$, respectively:  
\begin{align}
\label{nseq}
\frac{\partial v_{i}}{\partial t}+v_{j} \frac{\partial v_{i}}{\partial x_{j}}
&=-\frac{\partial p}{\partial x_{i}}+\nu \frac{\partial^2  v_{i}}{\partial x_{i}{^2}}+  \alpha gT \delta_{i3}\,,\\
\frac{\partial T}{\partial  t}+v_{j} \frac{\partial T}{\partial x_{j}},
&=\kappa \frac{\partial^2 T}{\partial x_{j}^2}+\kappa\frac{\partial^{2} \Theta}{\partial x_{j}^{2}}-v_{3}\frac{\partial \Theta}{\partial x_{3}}\,,
\label{pseq}
\end{align}

\noindent Here the horizontally averaged  temperature  $\Theta=\Theta(z)$ and $\partial_{j}v_{j}=0$.   It follows from equation (7) the balance:

$$\overline{{\cal E}}=-\alpha g\overline{v_{3}T}$$

\noindent stating that mean kinetic energy production by  temperature fluctuations  is balanced by the dissipation ratel.  Below,  this relation will be used for normalization, so that

$$e_{1}=\frac{\overline{{\cal E}}}{-\alpha g\overline{v_{3}T}}=1$$

\noindent and we will be interested in evaluation of all  moments $e_{n}=\frac{\overline{{\cal E}^{n}}}{(\alpha g)^{n}\overline{(v_{3}T)}^{n}}$.




\noindent According to (7), to understand small-scale features of a flow, we have to investigate temperature fluctuations  acting as a forcing term in the Navier-tokes equation (7).   Below, we use the theory of probability density (PDF) of temperature fluctuations in RB convection developed  in the nineties [11]-[12].  
 In the low-Rayleigh number linear  and weakly non-linear regimes, the following results, relevant for this study,  have been firmly established [13]-[17]. 

\noindent 1. \ At $Ra<1708$ the heat transfer is governed by conduction with heat flux $J=\kappa\frac{\partial \Theta}{\partial x_{3}}=\nu\frac{\partial \Theta}{\partial z}=const $ and ${\bf v}=0$;  \\ 
\noindent 2. \ First,  at $Ra\approx Ra_{cr}\approx 1700$, instability of a linear temperature profile with ${\bf v}=0$, typical of conduction, leads to formation of a "quasi-steady"  large-scale flow pattern  called rolls. \\ 
\noindent 3. \ Then,  in the interval $6\times 10^{4}-5\times 10^{6}$,  weak  fluctuations around this ordered  flow field lead to  the low - amplitude,   $O(Ra-Ra_{cr})\ll Ra_{cr})$,   velocity and temperature fluctuations.  In this range, according to  Krishnamurti [14] and Busse [15],  convection consists of ordered rolls with embedded small-scale  fluctuations  they call ``convection elements''.    It is important that,  while rolls are characterized by the length-scale $r\approx H$, the small-scale elements ``live'' on the scale $r\ll H$, independent on $H$. 
Quoting Busse [15]:  `'At moderate Prandtl numbers, turbulent convection at Rayleigh numbers of the order of $10^{5}-10^{7} $ exhibits the typical structure of relatively steady large-scale cells in which highly fluctuating (both in space and in time) small-scale convection elements are imbedded''.   Similar results have been reported in a detailed study of Castaing et.al. [13], showing  a few peaks on the heat transfer curves before  the rise of ``hard'' turbulence at $Ra\approx 4\times 10^{7}$. In a relatively recent paper, P. Tong  et. al. [17]  reported two {\bf competing} mechanisms of heat transfer originating  from the fluctuations   in a bulk of  convection  cell and plumes  produced by instabilities of  viscous sublayers. The most important  lesson from the existing experiments for  what follows  is emergence  of a few qualitatively different contributions to the heat transfer in a soft turbulence range of $Ra$ variation.\\
\noindent  To solve (7)-(8), following [10]-[12], the  domain of variation of both velocity and temperature fields,   can be  subdivided in two parts. a. \ The wall region  with thin  ($\eta<<H$)  velocity and temperature boundary layers (BL). Due to  the no-slip boundary conditions  ${\bf v}_{top}={\bf v}_{bot}=0$,   strong  wall shear leads to the boundary layer instability 
manifested in discrete bursts in the directions of  the bulk.  Similar mechanism of turbulence production in channel flows, responsible for the low Reynolds number Blasius scaling of the friction coefficient, was recently discussed in [16]. 
This leads to generation of velocity/temperature fluctuations in the bulk. \\
 \noindent b. \ Thus, we will study convection outside  boundary layers, using the  phenomenology of the  BL physics as an approximate {\bf boundary conditions}  for equations defined in the bulk.   In this domain turbulence can be assumed  isotropic and homogeneous. \\
 \noindent 4. \ In the high Reynolds number limit $Ra\gg Ra_{cr}$,  the flow becomes strongly non-linear and the notion of well-separated plumes invalid: due to strong interaction they loose their individuality in the bulk of the cell. 
This limit  is characterized by strong small-scale intermittency and anomalous scaling.


\section{Statistical ensemble.  Probability Density $P(X)$.  Low "Reynolds Number".}

\noindent Here we  consider an infinite fluid between two horizontal plates  separated by a gap $H$.
The integral scale of turbulence is $L\geq H$ and thus, the flow is generated in a huge number $N\approx {\cal L}/H\rightarrow\infty$  of independent  statistical realizations.  Therefore, one can use the theory of a passive scalar proposed by Sinai and Yakhot  [11]  and applied to the problem of  Benard convection  in Ref.[12]. 
 \\
Since in the field of  large-scale rolls, $\frac{\partial v_{3}}{\partial t}\approx V\frac{\partial v_{3}}{\partial z}$,   
   we define the  ``low - Reynolds number regime''  by the range where the non-linearity in (7) can be neglected,
\begin{eqnarray}
V\frac{\partial v_{3}}{\partial z}-\nu\frac{\partial^{2}v_{3}}{\partial x_{p}^{2}}=\alpha g T 
\end{eqnarray}

\noindent  {\bf plus no-slip boundary conditions on solid walls. } With ${\cal E}_{3}=\nu(\frac{\partial v_{3}}{\partial x_{p}})^{2}$:

$$\frac{{\cal E}_{3}}{\overline{{\cal E}_{3}}}+\frac{\frac{V}{2}\frac{\partial v_{3}^{2}}{\partial z}-\frac{\nu}{2}\frac{\partial^{2}}{\partial x^{2}_{p}}v_{3}^{2}}{\overline {\cal E}_{3}}
=  \frac{ v_{3} T}{\overline{v_{3}T}}$$

\noindent From the heat equation we have:

$$\frac{(\nabla T)^{2}}{\overline{(\nabla T)^{2})}}=\frac{ v_{3} T}{\overline{v_{3}T}}$$   

\noindent Defining  $\hat{\cal E}=\nu(\frac{\partial v_{3}}{\partial z})^{2}$, gives:

$$\hat{\cal E}-\frac{2\nu^{2}}{V}\frac{\partial v_{3}}{\partial z}\frac{\partial^{2}v_{3}}{\partial x_{p}^{2}}+\frac{\nu^{3}}{V^{2}}(\frac{\partial^{2} v_{3}}{\partial x_{p}^{2}})^{2}=\nu (\frac{g\alpha}{V})^{2}T^{2}
$$

Based on  the theory  [11]-[12]  supported by experimental data [13]-[[15], [17], ]we conclude that there exist two mechanisms of  dissipation  of kinetic energy  
${\cal E}_{3,1} \propto  T^{2}$ and ${\cal E}_{3,2}\approx b|T|$, 
with   continuous  $O(T^{2})$ contribution coming from the  ``convection elements''  and the $O(T)$ one from the discrete plumes arising from the BL 
instability at $Ra\geq 1700 \eta^{bl}/H$ with the typical  rising velocity $V$. Obtaining this estimate we relied on the concept ``marginally stable'' boundary layer introduced by Malkus [18] and discussed  Castaing et.al [13]. \\

\noindent   We are interested in probability density $P(X)$  in the limit of small $X=T/T_{rms}$.   This range includes heat conduction regime,  formation of  weakly fluctuating rolls and discrete plumes coming boundary layers.  As in Refs. [10] -[12], it is assumed that in the central part of the cell the fluid is well mixed and turbulence there can be assumed homogeneous and isotropic.  Following [11], [12]
 multiplying (8) by $T^{2n-1}$ gives:

$$-(2n-1)\overline{T^{2n-2}(\nabla T)^{2}}=\overline{T^{2n-1}v_{3}}\frac{\partial \Theta}{\partial z}$$

\noindent With $X^{2}=T^{2}/\overline{T^{2}}$, $Y^{2}=(\nabla T)^{2}/\overline{(\nabla T)^{2}}$ and $W=v_{3}T/\overline{v_{3}T}$.  
These equations can be rewritten:

$$(2n-1)\overline{X^{2n-2}Y^{2}}=\overline{X^{2n-2}W}$$

\noindent and introducing conditional means gives [10]-[12]:

$$(2n-1)\int X^{2n-2}r_{1}(X)P(X)dX=\int X^{2n-2}r_{3}(X)P(X)dX$$ 

\noindent where 

$$r_{1}(X)=\frac{\int Y^{2}(x)\delta(X(x)-X)dx}{\int \delta(X(x)-X)dx}$$

\noindent and

$$r_{3}(X)=\frac{X}{\overline{v_{3}X}} \frac{\int v_{3}(x)\delta(X(x)-X)dx}{\int \delta(X(x)-X)dx}$$

\noindent $r_{1}(X)$  and $r_{3}(X)$ are conditional expectation values of temperature dissipation and production  rates for fixed 
magnitude of dimensional temperature $X$. After simple manipulations one obtains a formal expression for 
 probability density $P(X)$ [10]-[12]:

$$P(X)=\frac{C}{r_{1}(X)}\exp\big[-\int_{0}^{X}\frac{r_{3}(u)du}{ur_{1}(u)}\big ]$$

\noindent or 
\begin{eqnarray}
P(X)=\frac{C}{r_{1}(X)}\exp\big[-\int_{0}^{X}\frac{uv_{3}(u)du}{ur_{1}(u)}\big ]
\end{eqnarray}

\noindent We can  evaluate  this expression in the limit $X\rightarrow 0$.  First, according to [11]-[12],  positive definite conditional dissipation rate 

 $$r_{1}(X)\approx \alpha +\beta X^{2}=\alpha(1+\frac{\beta}{\alpha}X^{2}) $$

\noindent  Since positive temperature fluctuations (blobs of hotter fluid)  are carried by positive velocity fluctuations $v_{3}$,  we conclude that $v_{3}(T)\approx -v_{3}(-T)$.    

\noindent As $Ra-Ra_{cr}\rightarrow 0$, the fluctuations of the large-scale rolls, called ``convection elements''     are very weak,  lacking  any typical velocity scale.  Therefore, in this limit  by the symmetry: $v_{3}(X)\propto X$. 
At larger Rayleigh number the instability of viscous sublayers leads to plumes emitted with a typical velocity $V=yV_{p}$ where we introduce an artificial Reynolds number $y\propto Re-Re(p)$ with  $Re(p)$  denoting  the Reynolds number of first instability of boundary layer (manifested in peaks  in  a heat flux  curve) resulting in  weak discrete bursts.  This means that in this theory $y\geq 0$ and 
the conditionally averaged velocity can be written as:

$$\frac{v_{3}(X)}{v_{rms}}\approx  \gamma X+2yV_{p}/v_{rms} \approx \gamma X+2\kappa y$$

\noindent where $V_{p}/v_{rms}=O(1)$.  We can see that when $y=0$, the resulting Gaussian flow is dominated by the weak small-scale elements. Substituting all this into (10) gives:

$$P(X,y)=\frac{C(y)}{(1+\frac{\beta}{\alpha} X^{2})}\exp[\big[-\int_{0}^{X}\frac{\gamma u+2\kappa y}{\alpha(1+\frac{\beta}{\alpha} u^{2})}du \big ]$$ 

\noindent and the probability density of temperature fluctuations in the central part of convection cell with $\alpha=\kappa=1$ is:
\begin{eqnarray}
P(X,y)=\frac{C(y)}{(1+ \frac{\beta}{\gamma}X^{2})^{1+\frac{\gamma}{2 \beta}}} exp(-2 y \arctan(\sqrt{ \beta}X)\equiv \nonumber\\
C(y)\Pi(X,y)
\end{eqnarray}

\noindent  with    $C(y)=1/ 2\int_{0}^{\infty}\Pi(X,y) dX $ 
and  $\beta\approx 1.4$ estimated in [12].  As $y\rightarrow 0$,  this expression gives Gaussian with the half-width $\delta\approx \sqrt{\gamma/\beta}$.   It has been found  in Ref.[12] that although the derivation is, strictly speaking, valid for $\frac{\beta}{\gamma}X^{2}\rightarrow 0$, the result agrees very well with numerical simulations in a much broader interval.  An interesting feature of this expression is the dependence of the  PDF on Reynolds number $y$. This is the consequence of a qualitative transition happening in the flow  $y>0$. The behavior of the PDF as a function of ``Reynolds number'' $y$ is shown on Fig.1.

\subsection{Moments of dissipation rate. Low-Re regime.}

\noindent Based on the above derivation (also see Ref.[12]),   the conditional mean of kinetic energy dissipation rate is approximated by the expression:

\begin{eqnarray}
\frac{\cal{E}}{\overline{\cal E}}\approx yX+X^{2}
\end{eqnarray}

\noindent  and thus, the normalized  moments of the dissipation rate are calculated readily

$$   e_{n}(y)= \frac{\int_{0}^{\infty} (yX+X^{2})^{2n})P(X,y)dX}{(\int_{0}^{\infty} (yX+X^{2})^{2})P(X,y)dX)^{n}}$$

\begin{figure}
\includegraphics[scale = 0.4]{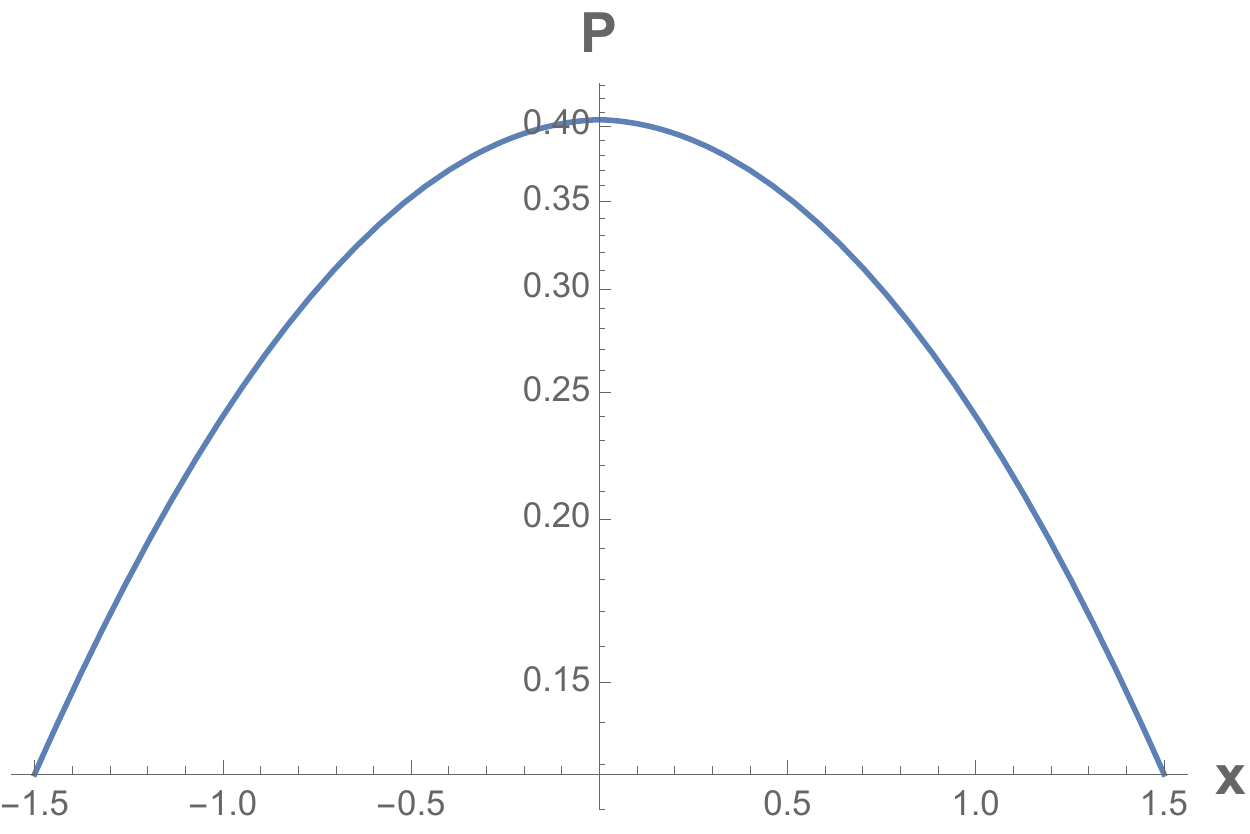}
\includegraphics[scale = 0.4]{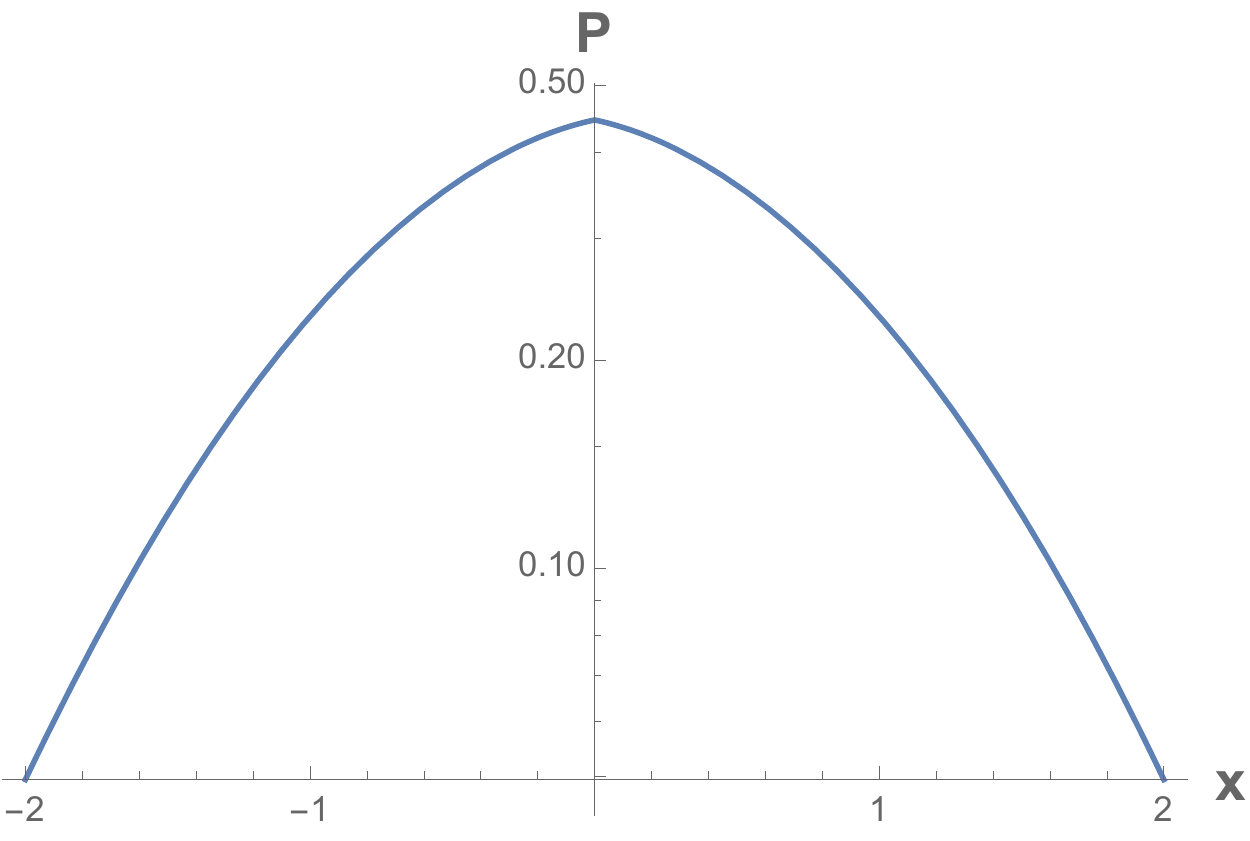}
\includegraphics[scale = 0.4]{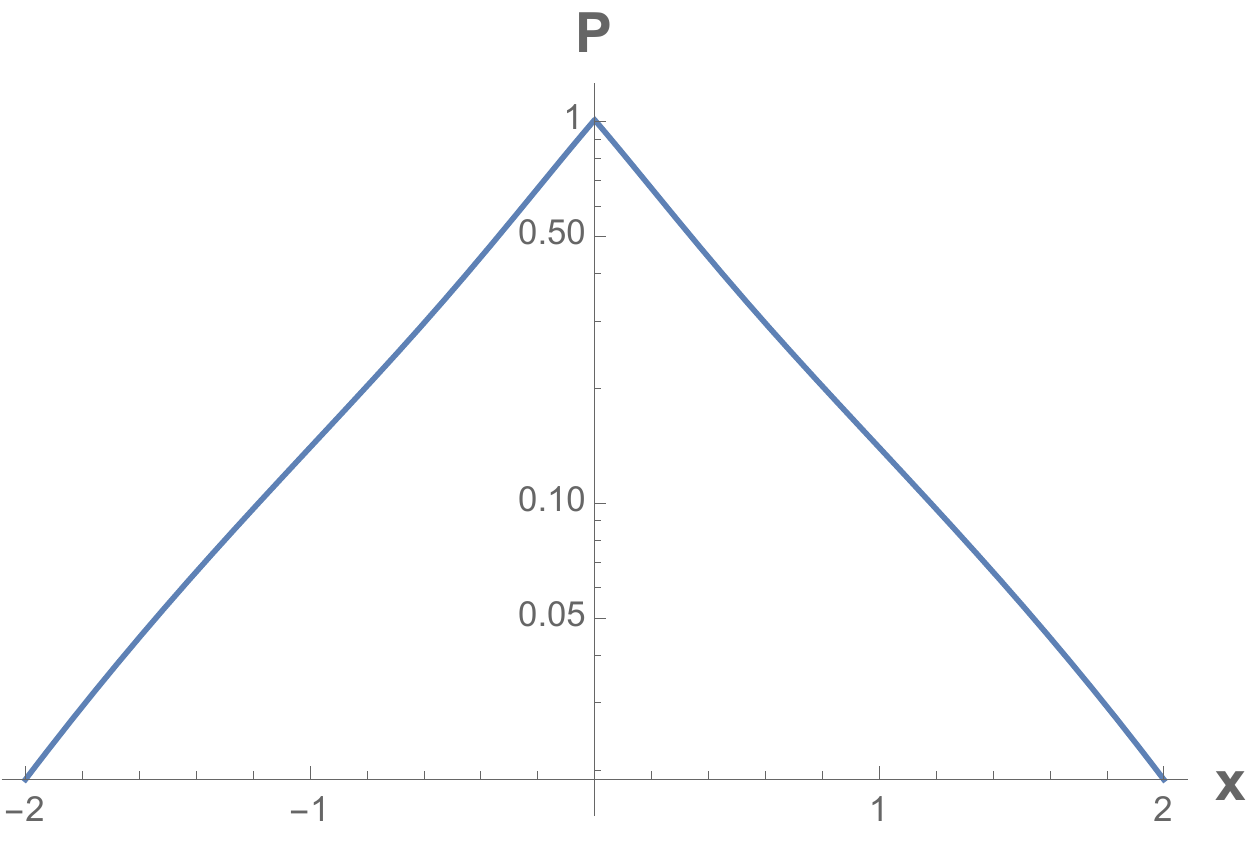}
\caption{ Probability  densities  of normalized dissipation rate $e={\cal E}/\overline{\cal E}$  vs   "Reynolds number" $y$.\\ 
Top panel: $y=0.01$. Middle: $y=0.1$. Bottom: $y=1.$. In all  cases $\beta=1.4$  as estimated in [12]-[13].  }
\label{fig1}
\end{figure}
 
 \begin{figure}
\includegraphics[scale = 0.4]{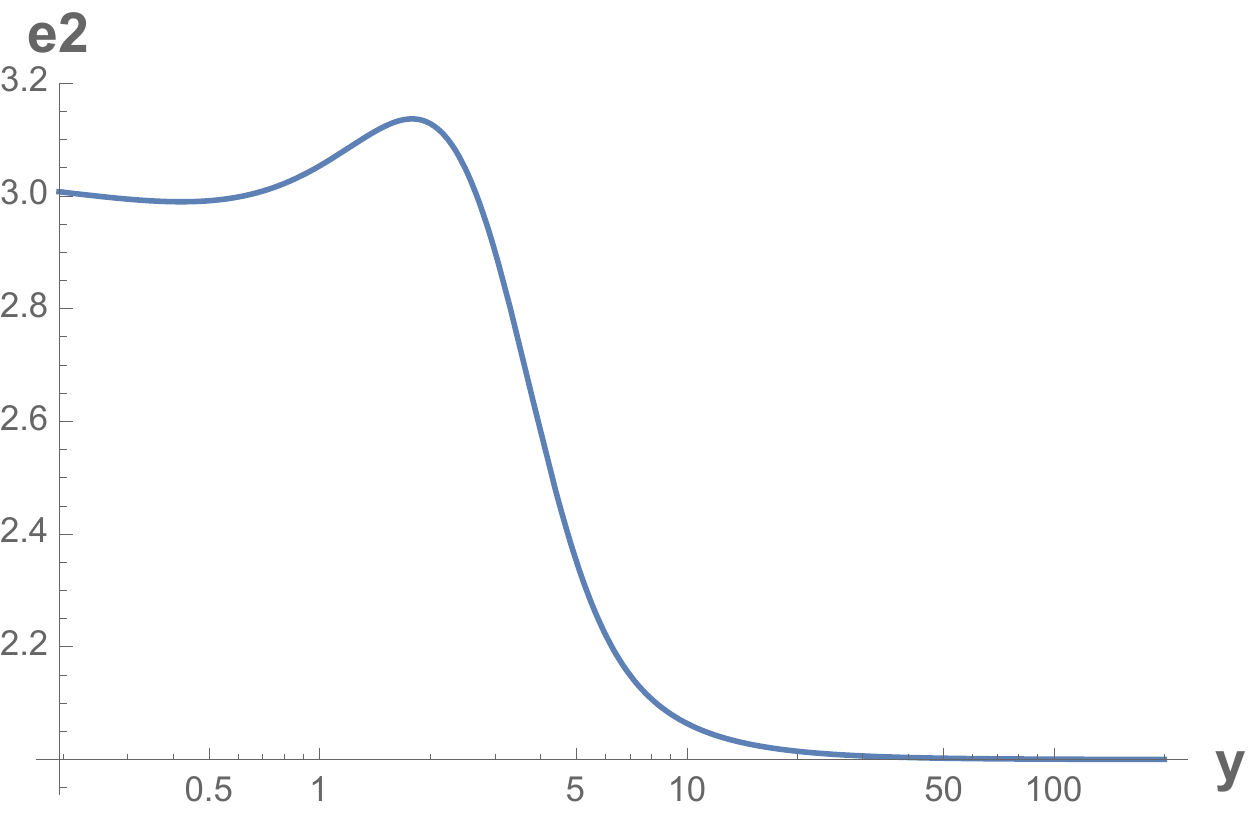}
\includegraphics[scale = 0.4]{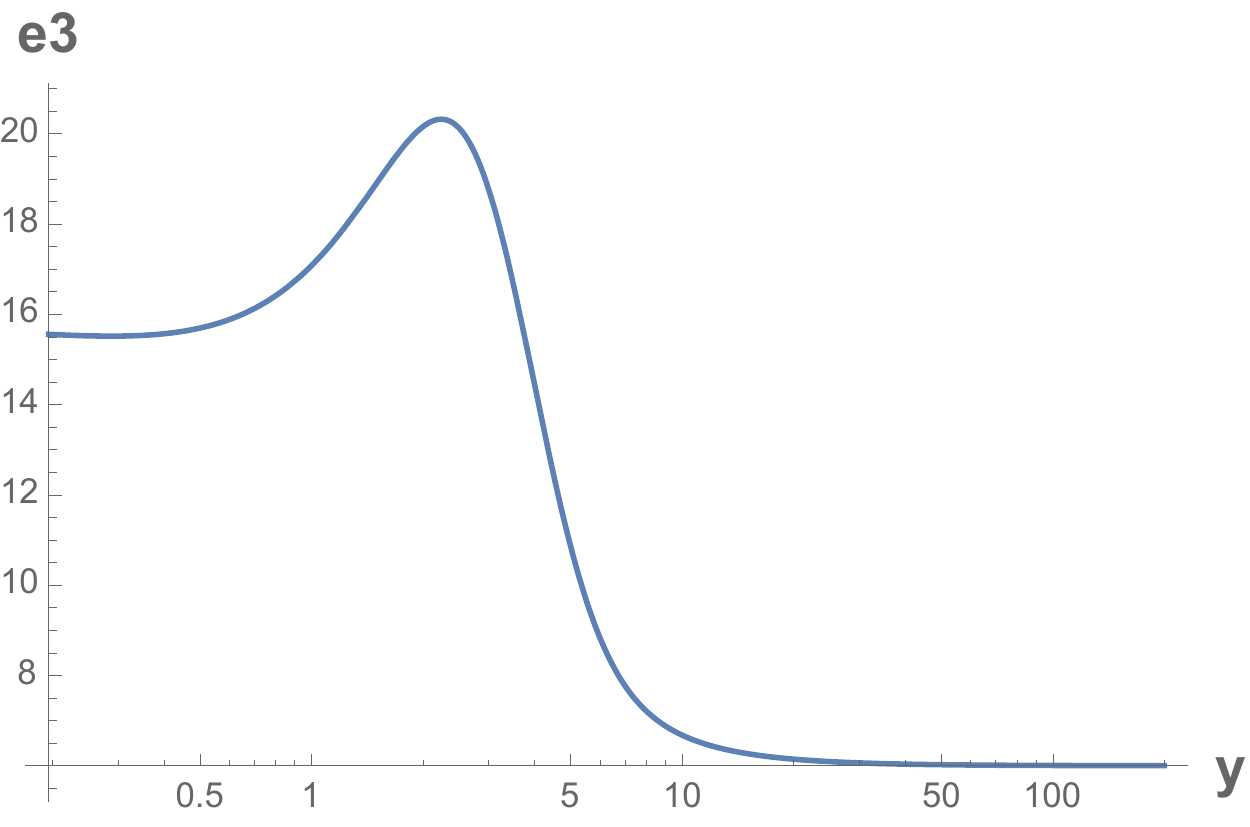}
\includegraphics[scale = 0.4]{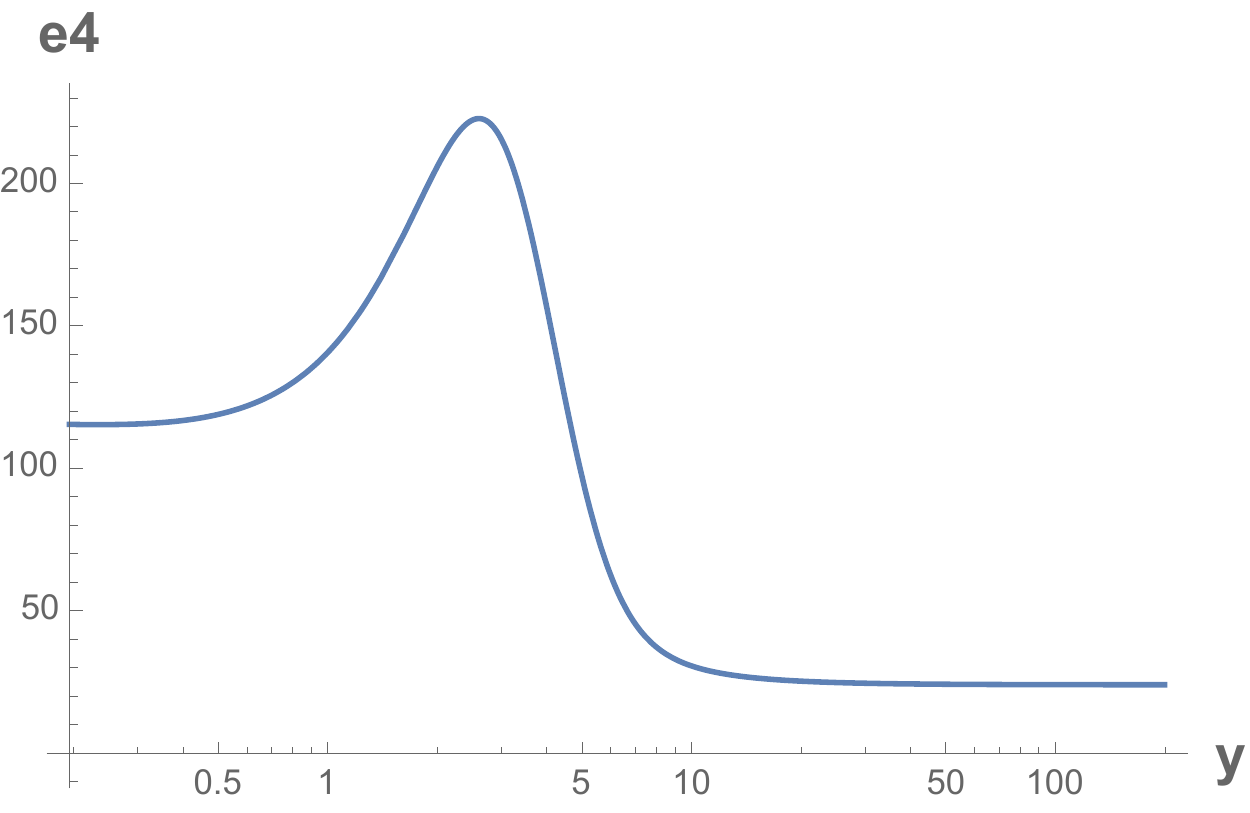}
\caption{Nonmonotonic low- Reynolds number behavior of normalized moments of kinetic energy dissipation rate $e_n=\frac{\overline{{\cal E}^{n}}}{\overline{\cal E}^{n}}$  as a function of artificial " Reynolds number"  $y\propto Re-Re(p)$ , first observed in Ref.[10].   At $y\rightarrow 0$,  all $e_{n}(y)\approx (2n-1)!!$ indicating Gaussian statistics.  With increase of $y$, one can see transition to a state dominated by weak structures   and close-to-exponential probability densities.  The range $Re >>Re_{n}^{tr}$ corresponds to strong coupling where the quasi-linear approximation breaks down.}
\label{fig2} 
\end{figure}

\begin{figure}
\includegraphics[scale = 0.4]{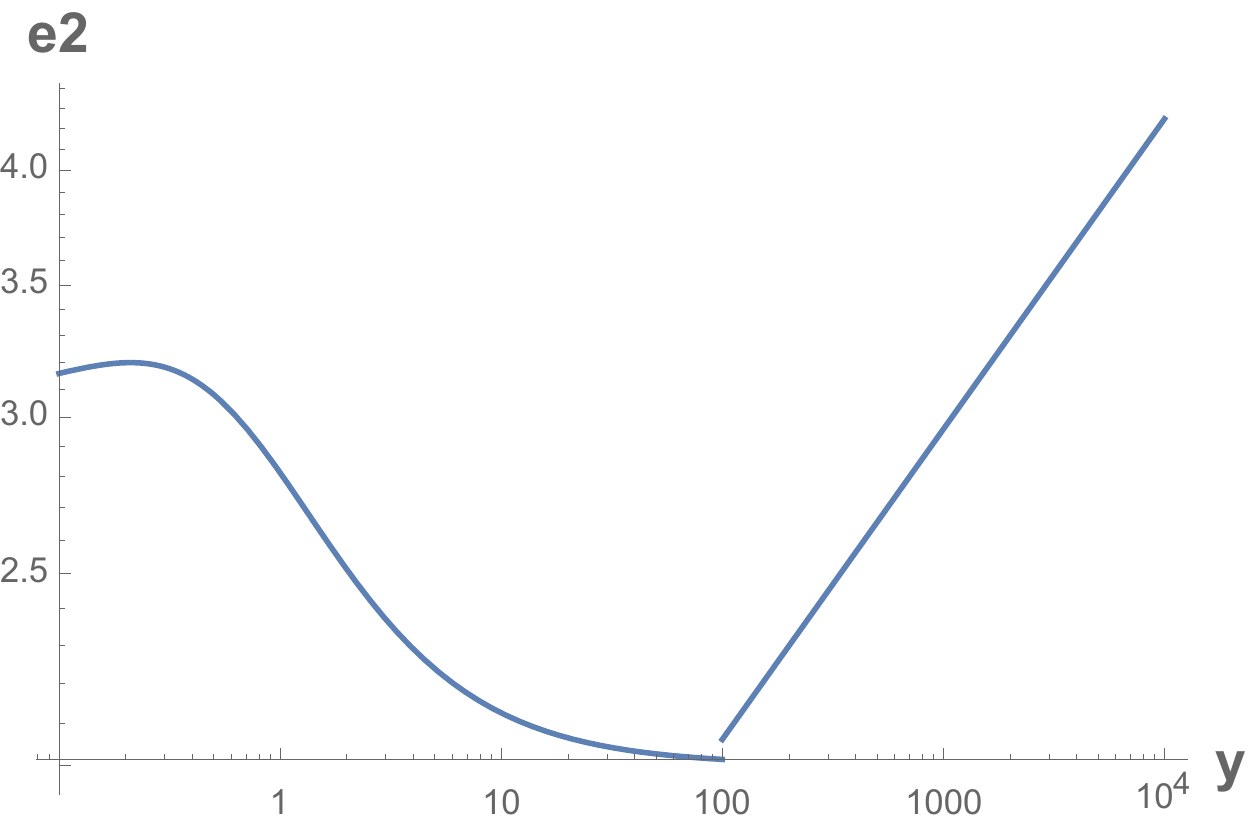}
\includegraphics[scale = 0.4]{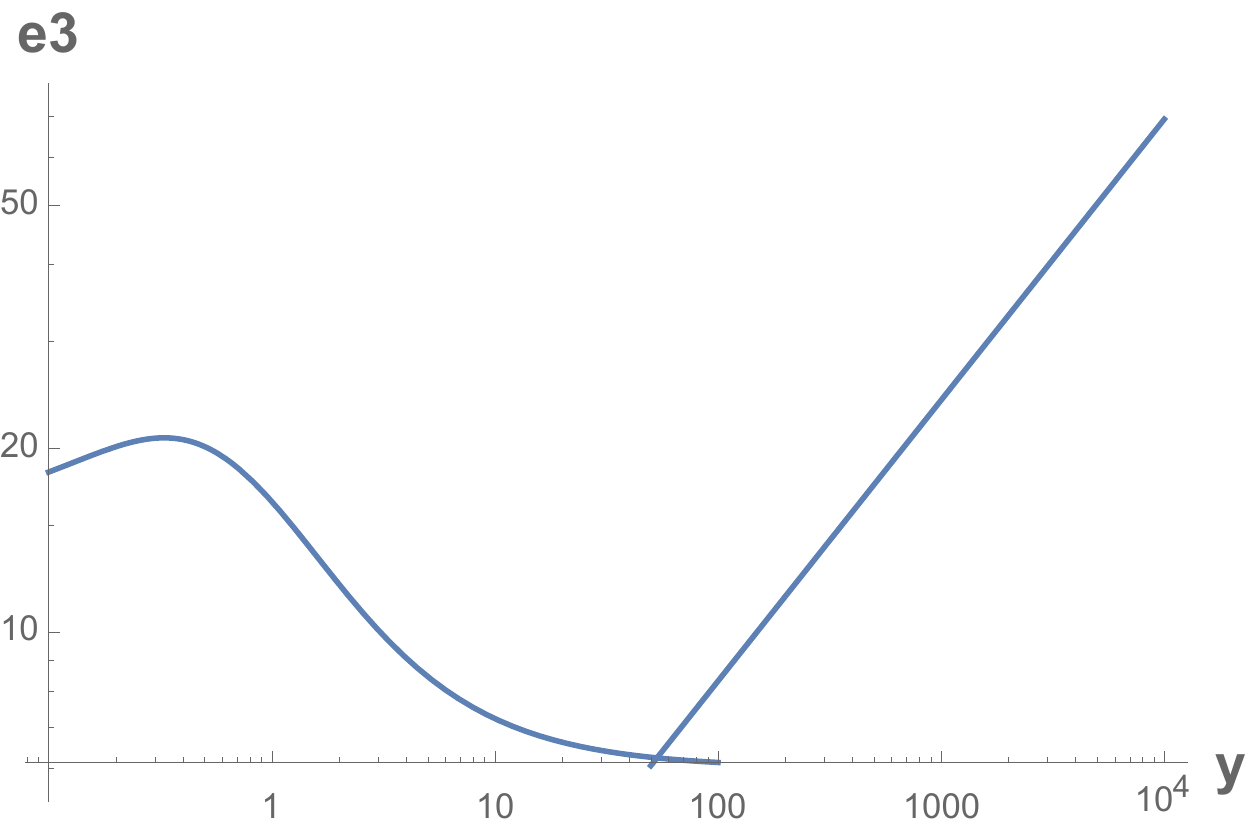}
\includegraphics[scale = 0.4]{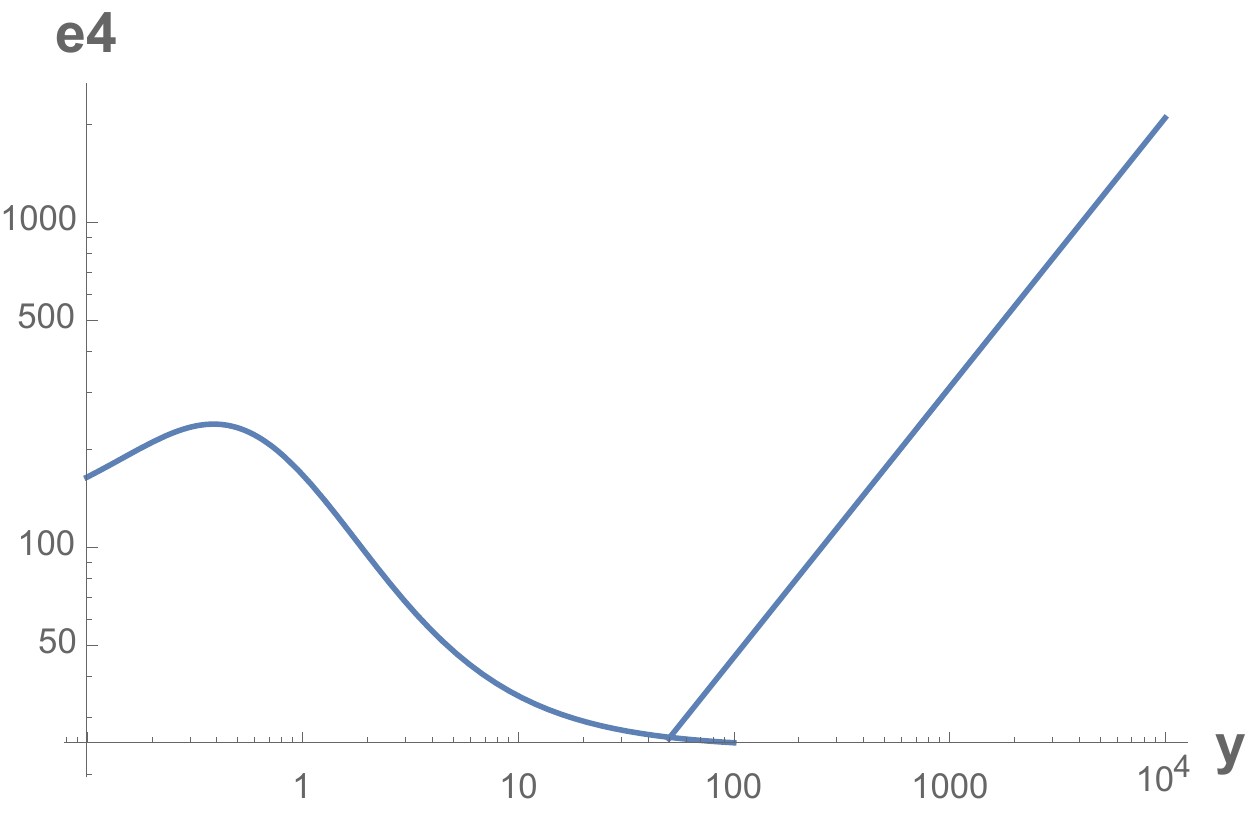}
\caption{ Analytic procedure for evaluation of moments of dissipation rate in the entire Reynolds number range $0 \leq y  \leq \infty$.     Given the low-Reynolds number 
moments $e_{n}(y)$  (Fig.3)   and    transitional Reynolds numbers $Re_{2}^{tr}\approx 100$,    the  exponents  $d_{n}(y)$ and $\rho_{2n}(y)=d_{n}(y)+n$ for the moments  $e_{n}(y)$ and $M_{2n}(y)$, vali
in   strongly non-linear,  anomalous, regime.  $Re_{n}^{tr}\leq  Re< \infty$ are calculated from (14).   The two limiting curves match at $Re=Re_{n}^{tr}$.   On this graph: $Re\rightarrow \infty$, $e_{2}\propto Re^{0.157}$, $e_{3}\propto Re^{0.46}$ and $e_{4}\propto Re^{0.83}$.  }
\label{fig3}
\end{figure}

\begin{figure}
\includegraphics[scale = 0.6]{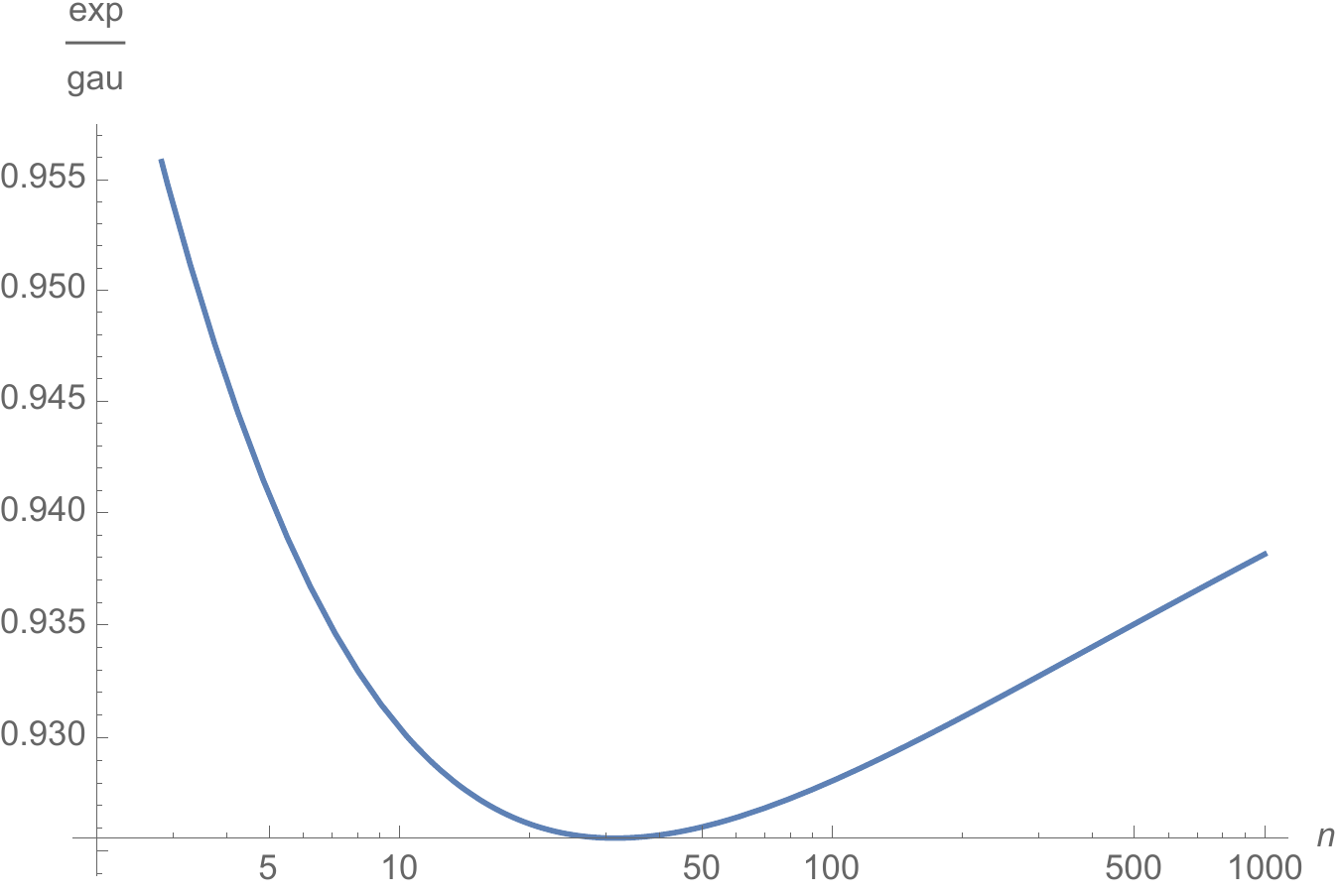}
\caption{ Ratio of scaling exponents  $\rho_{2n}=d_{n}+n$ in flows driven by exponential and gaussian random forces  vs $n$. . 
}
\label{fig3}
\end{figure} 

\noindent {\bf This expression  is valid when Reynolds number is so small that the non-linearity  in (7)  can be neglected, but large enough   to allow for the  relatively weak 
boundary layer instability  leading to isolated (discrete)  plumes. This mechanism   is similar  to the one considered in [16]  responsible    to the  intermediate Blasius scaling in a channel flow.}\\
In the interval  $y\ll X$,  $yX<X^{2}$ and the probability density $P(X)$  is close to the Gaussian with the first few low-order moments 
$e_{n}\propto \overline{X^{2n}}\approx (2n-1)!!$.  

\noindent It is interesting  that the expression (12)  with $y\propto Re-Re^{tr}$,  reflects two competing mechanisms experimentally observed by Tong et.al.  [17].  Indeed, when 
$y\rightarrow 0$, the scale-lacking-excitations dominate the Gaussian PDF.   
One can also see, that as the ``Reynolds number"  $y$ grows,  due to  appearance of discrete plumes $yX>X^{2}$,  the PDF  (11) varies to close- to -exponential which is an immediate precursor to  anomalous scaling and intermittency. 
The  smooth transition from $e_{n}=(2n-1)!!$ to $e_{n}=n!$, experimentally  and numerically  observed in  Refs.[13] and [10],  respectively, is shown on Figs. 2-3.

  \subsection{Matching condition:   Strong turbulence, Intermittency in Benard convection}
  
\noindent  It follows from the theory developed in [4] (also see Section I) that,  to describe strongly non-linear  limit of turbulent fluid, $Re\rightarrow \infty$,  
one has to understand fluid behavior in the weakly non-linear range $0\leq Re\leq R^{tr}_{n}\approx 120$.   It is a matching 
of low and high -Reynolds -number asymptotic solutions  gives an equation for the amplitudes and anomalous exponents in the strong turbulence interval $Re_{n}^{tr}\leq Re < \infty$. 
 Direct transition from from ``normal'' to ``anomalous''  scaling in the   Gaussian-force-driven fluid, described in Ref.[3]-[4], [10]  and in Section II of this paper,  is relatively simple:  the 
derivative moments are equal to  $M_{2n}=(2n-1)!!$ in the low-Re range $0\leq  Re_{n}^{tr}\approx 100-120$ or $R^{tr}_{\lambda,n}\approx 8.91$.

If at $Re\ll Re_{n}^{tr}$   the moments are  $e_{n}=(2n-1)!!$ or $e_{n}=n!$  as in Gaussian or exponential cases, respectively, one has to understand relations between these states as a function of Reynolds number. It is promising that the  low-Re range behavior of  a flow can be addressed numerically using direct numerical simulations. In a general case of the Reynolds number dependent moments, the exponents $d_{n}(y)$ are found from the equation:

\begin{eqnarray} 
e_{n}(y)= \overline{({\cal E}/\overline{\cal E})^{n}}=C^{d_{n}(y)}(\hat {R}^{tr}_{\lambda,n})^{\frac{nd_{n}(y)}{d_{n}(y)+\frac{3n}{2}}}
\end{eqnarray}
 
\noindent where $e_{n}(y)$ are found from Fig.3. The result is. 

\begin{eqnarray}
d_{n}(y)=-\frac{1}{2}[n(\frac{2.19}{\ln C}+\frac{3}{2})-\frac{\ln e_{2n}}{\ln C}]+ \nonumber \\ 
\sqrt{\frac{1}{4} [n(\frac{2.19}{\ln C}+\frac{3}{2})-\frac{\ln e_{2n}}{\ln C}]^{2} +\frac{3}{2}n\frac{\ln e_{2n} }{\ln C}}
\end{eqnarray}


\section{Summary and discussion. Universality.}

The direct transition from a   Gaussian flow  ( $R_{\lambda}^{tr}\leq 8.91$), was investigated both theoretically and numerically  in  Ref[.3]-[4].  The results, based on transitional Reynolds number 
$R_{\lambda,2}^{tr}=8.91$, and the amplitude $C\approx 90-100$,  are presented on Fig.1. 
It is clear from Fig.2 that in the case of RB convection the low-Re dynamics are  much more involved and in the limit $y\propto Re-Re_{cr}\rightarrow 0$, the flow is indeed close-to-Gaussian and can be treated using the results  of Sec.I. However, to obtain the moments at  a transitional Reynolds number, precursor to anomalous scaling, one has to understand  fluid dynamics at  the moderate Reynolds numbers $Re\leq Re_{n}^{tr}\approx 100$ or $R_{\lambda}\approx 8.91$.  The main question is: {\bf how universal this number is?}\\

\noindent The universality of the Reynolds number based on ``turbulent'' viscosity $R_{\lambda, T} \approx 10.0$,  derived from dynamic  Renormalization Group  [5]-[8], widely used  in engineering [9], is known  for many years. In fact, it is the basis of the so-called 
${\cal K}-{\cal E}$ modeling (see Section I) and Ref.[9].  Numerical and experimental data on  flows  past  the cylinder, decaying turbulence and even flow past various 
industrial applications  like cars,  gave   for the  Reynolds number based on ``turbulent viscosity'' $R_{\lambda}^{T}\approx 9.0-11.0$.
In Ref.[5]  the  transition to anomalous scaling $R_{\lambda}^{tr}\approx 9.0$ has been first reported in the DNS of the Navier- Stokes equations on a periodic domain driven by a force ${\bf f}\approx \alpha {\bf v}$ defined  at the large scales with $\frac{2\pi}{L}=k\approx 1-2$, completely different from the one discussed in Refs.[3]-[4].  Possible universality of this number 
may be not too startling. Indeed,  while in open, far from equilibrium,  system 
 the ``bare''  $Re_{cr}$  of the first instability  of a laminar pattern may  vary in a  broad interval,  the ``dressed'' one,  characterizing transition from ``normal'' to anomalous dimensions (intermittency) can be  fixed at $R^{tr}_{\lambda}\approx 10$.    The possible universality of transitional $R_{\lambda,tr}\approx 9.0$ may have interesting implications. Anomalous scaling is usually related to coherent structures appearing in   a coherence-lacking background random flow.   If this is so, then it is not impossible that universality of $R^{tr}_{\lambda}$ may indicate universal, flow-independent, structures responsible for transition to strong turbulence. In the future publication 
 The variation of flow geometry at $R^{tr}_{\lambda}\geq  9.0$ has recently been reported by Das and Girimaji [19] in homogeneous and isotropic turbulence driven by a random force.  How universal the effect is remains an open and interesting question. We expect the "magic number" $R_{\lambda}^{tr}\approx 8.91$ to be related to Feigenbaum  numbers describing transition in terms of period-doubling mechanism.\\

\noindent To study the role of the forcing statistics we, assuming for the sake of   argument,   universality of constants $R_{\lambda,2}^{tr}\approx 9.0$,  and $C\approx 90-100$,  evaluated the exponents $d_{n}$  from  the expression (14).  The ratio of exponents $\rho_{2n}=d_{n}+n$ in the flows driven by   gaussian and exponential  forces, respectively,     is plotted on Fig.5,  for $y=4.5$ , in the huge, not experimentally  realizable interval $2\leq n \leq 1000$. One can see the ratio varying in the range   $0.925\leq \frac{exp}{gau}\leq 0.955$,  which, though quite close to unity,  may indicate existence of universality classes reflecting mechanisms driving turbulence flow.

\noindent To conclude the paper we would like to pose a question which can  readily  be resolved in future numerical and physical experiments: 
how general is the passage  to turbulence,  described in this paper,  in a typical wall flow where the randomness-generating  bulk and wall-layer instabilities  often  coexist ?\\

\noindent Given the results   of Ref.[16], this generality may not be impossible. The role of weak-to-strong turbulence transition in chemical kinetics, combustion and mixing in high-Reynolds number  fluids may be of importance in various, at present not explained, processes.

\section*{Acknowledgements.} 
 The ideas leading to this paper were discussed in a recent  Turbulence Workshop   (Texas A\&M University, August 30-31, 2018). I am grateful to 
 J.Schumacher, D.Donzis, K.R.Sreenivasan and S.Girimaji for  discussions of various aspects of the problem. DNS of RB convection 
 performed by J.Schumacher and his team  served  as a first impulse which resulted in  this  paper.  Many thanks are due to A.Polyakov who brought  my attention to applications of a somewhat different matching condition for  
 non-perturbative evaluation of anomalies in QCD [20]. Also,  I appreciate the input of Drs. Chen and Staroselsky   
 of EXA Corporation for 
 sharing  a lot of data on ``turbulent''  Reynolds numbers in various applications.

\end{document}